\documentclass[11pt,a4paper]{article}
\usepackage{amsmath,amssymb}
\usepackage{epsfig,graphicx}

\newtheorem {theorem}{Theorem}[section]

\begin{document}
\title{\bf Recent Developments in the Skyrme Model}
\author{Steffen Krusch$^a$\footnote{{\bf e-mail}: 
S.Krusch@kent.ac.uk},
Mark Roberts$^b$\footnote{{\bf e-mail}: M.Roberts@surrey.ac.uk}
\\
$^a$\small{\em Institute of Mathematics, University of Kent,} 
\small{\em Canterbury CT2 7NF, United Kingdom}\\
$^b$\small{\em Department of Mathematics, University of Surrey,} 
\small{\em Guildford, GU2 7XH, Surrey, United Kingdom}
}
\date{September 2008}
\maketitle

\begin{abstract}
\noindent
In this talk, we describe recent developments in the Skyrme
model. Our main focus is on discussing various effects which need to
be taken into account, when calculating the properties of light atomic
nuclei in the Skyrme model. We argue that an important step is to
understand ``spinning Skyrmions'' and discuss the theory of relative
equilibria in this context.
\end{abstract}

\section{Introduction}

The Skyrme model is a classical field theory modelling the strong
interaction between atomic nuclei \cite{Skyrme:1961vq}. It has to be 
quantized in order to compare it to nuclear physics.
Since Adkins et al. \cite{Adkins:1983ya,Adkins:1984hy}
quantized the translational and rotational zero-modes of the
$B=1$, a lot of progress has been
made in understanding both the classical solutions of the Skyrme model
and the quantization of Skyrmions. The rational map ansatz 
\cite{Houghton:1998kg} 
gives a very successful approximation to Skyrmions when the pion mass
is zero and helped finding all the Skyrmions up to baryon number
$B=22$, \cite{Battye:2001qn}. It was shown in
\cite{Finkelstein:1968hy} that solitons in
scalar field theories can consistently be quantized as fermions
provided that the fundamental group of configuration space has a
${\mathbb Z}_2$ subgroup generated by a loop in which two identical
solitons are exchanged. Noting that such loops arise from the
symmetries of classical Skyrme configurations lead to the calculation
of the quantum 
ground states up to $B=22,$ \cite{Irwin:1998bs, Krusch:2002by}.  
Currently, there is a lot of interest in understanding light atomic
nuclei with the aid of the Skyrme model. Now, the focus has shifted to
understanding excited states and making quantitative predictions about
their excitation energies. There are a number of issues which need to be
taken into account. The Skyrme model only has three
parameters. Traditionally, the value of
\cite{Adkins:1983ya,Adkins:1984hy} have been used, but there are now a
number of separate arguments why different values of the Skyrme
parameters should be used instead, \cite{Battye:2005nx, Manton:2006tq}. 
It was also discovered that the pion mass has a big influence on the
shape of Skyrmions, 
\cite{Battye:2004rw, Battye:2006tb, Houghton:2006ti}, in particular for $B>7$. 
Finally, it is important to take into account that Skyrmions change
their shape when they rotate, see \cite{Battye:2005nx,
  Houghton:2005iu} for a discussion of $B=1$. In this talk, we
describe how the theory of relative equilibria \cite{Montaldi:1999,
  Kozin:1999, Kozin:2000}
may shed light on spinning Skyrmions.

The talk is organized as follows. Section \ref{SkyrmeModel} gives a
brief introduction to the Skyrme model, the rational map ansatz and
the quantization of Skyrmions. In Section \ref{Rel}, we discuss the
theory of relative equilibria. The final section describes preliminary
results of applying the theory of relative equilibria to the Skyrme
model and gives an outlook to future work.

\section{The Skyrme Model}
\label{SkyrmeModel}

The Skyrme model is a nonlinear theory of pions which models the 
strong interaction between atomic nuclei. It is defined in 
terms of a field $U(t,\mathbf{x}) \in SU(2)$. The Skyrme Lagrangian is given by
\begin{equation}
\label{Lagrangian}
L=\int \left( -\frac{1}{2}\,\hbox{Tr}\,(R_\mu R^\mu) 
+ \frac{1}{16}\,\hbox{Tr}\,([R_\mu,R_\nu][R^\mu,R^\nu]) 
+ m^2\,\hbox{Tr}\,(U - 1_2) \right) d^3 x \,,
\end{equation}
where $R_\mu = (\partial_\mu U)U^{\dag}$, and $m$ is proportional to
the pion mass, \cite{Battye:2004rw}.
The Skyrme Lagrangian has localized finite energy solutions which 
behave like particles and are known as Skyrmions.

The Lagrangian is written in ``geometric units'' in which length is
measured in units of $2/e f_\pi$ and energy in units of
$f_\pi/4e$. The parameters $f_\pi$ and $e$ are known as the pion decay
constant and the Skyrme constant, respectively, and will be discussed
in more detail later.

In order to have finite energy, Skyrme fields have to take a constant
  value, $U(|{\bf x}| = \infty) = 1$, at infinity. Therefore,
from a topological point of view the Skyrme field $U$ at a fixed time can
be regarded as a map $U: S^3 \to S^3$, and such maps are characterized
by an integer-valued winding number. 
This topological charge is interpreted as the baryon number,
which for our purposes can be thought of as the number of protons and
neutrons.

The Skyrme Lagrangian is invariant under the Poincar\'e group of
Lorentz transformations and translations in space, together with
rotations in target space. There is also a discrete symmetry
\begin{equation}
{\cal P}: U({\bf x}) 
\mapsto U^\dagger (-{\bf x})
\end{equation}
known as parity. In the following, we will describe static
solutions. We will ignore translations and focus on the symmetry group 
$SO(3) \times SO(3):$ rotations in 
space and rotations in 
target space (known as isorotations). As usual, 
rotations are given by ${\bf x} \mapsto R {\bf x}$ where $R^T R = 1$,
and  isorotations by $U \mapsto AUA^\dag$, where $A \in SU(2)$.
These symmetries will play a very important 
role for the calculation of quantum states in the Skyrme model.

\subsection{The Rational Map Ansatz}
\label{Ratmap}
In this section, we describe the rational map ansatz \cite{Houghton:1998kg} 
which is a very successful approximation to minimal energy Skyrme 
configurations. The most convenient way for obtaining the explicit formula 
is the geometric approach of Manton \cite{Manton:1987xt, Krusch:2000gb}. 

\begin{center}
\begin{figure}
\begin{center}
\includegraphics[width = 10cm]{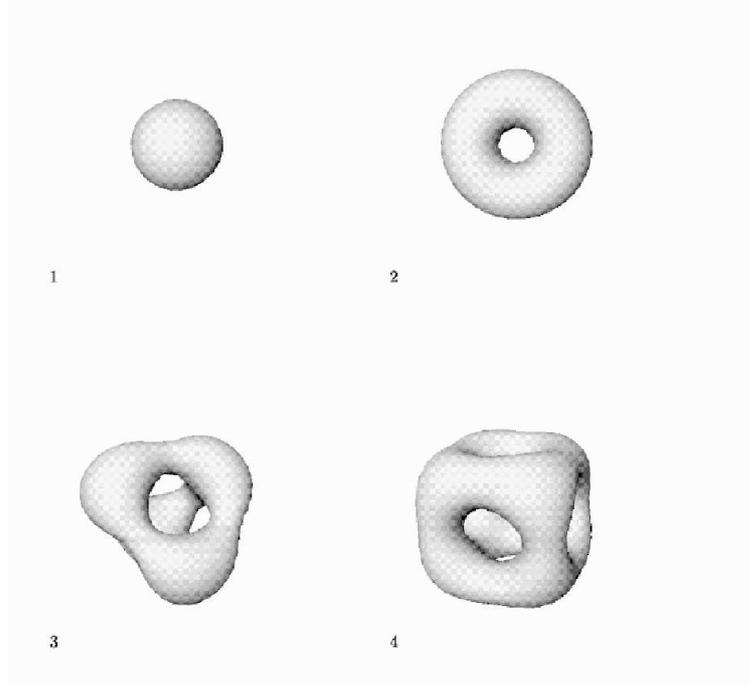}
\caption{Energy density level sets for $B=1,\dots, 4$.
(Figure taken from \cite{Battye:1997nt}).\label{F1}}
\end{center}
\end{figure}
\end{center}

The main idea is to use rational maps, i.e. holomorphic maps between
two-spheres in space and target space. In complex stereographic coordinates, 
$$
S^2 \rightarrow {\mathbb C} \cup \infty: (\theta,\phi) \mapsto z = 
\hbox{tan}\left(\frac{\theta}{2}\right){\rm e}^{i\phi}\,, 
$$
rational maps can
simply be written as ratios of polynomials $p(z)$ and $q(z)$,
$$
R(z) = \frac{p(z)}{q(z)}.
$$
The inverse map for the stereographic projection is
$$
\mathbf{n}_z = \frac{1}{{1 + |z|^2}}
\left(
\begin{array}{c} z + \bar{z}\\ i(\bar{z}-z)\\ 1 -
|z|^2
\end{array}
\right).
$$
The ansatz for the Skyrme field is then given by
$$
U(r,z) = \hbox{exp}\left(if(r)\,\mathbf{n}_{R(z)} \cdot
\boldsymbol{\tau}\right), 
$$
where $\boldsymbol{\tau}$ denotes the Pauli matrices.
This approximation leads to various simplifications. Angular and
radial integrals decouple, and the angular integrals only depend on a
finite number of parameters. The baryon number is given by
$$
B = - \frac{1}{2 \pi^2} \int f' \sin^2 f \left(
\frac{1 +
| z|^2 }{ 1 + |R|^2} \left\vert\frac{dR }{ dz }\right\vert
\right)^2\,\frac{2i\,dz\,d\bar{z}}{(1+|z|^2)^2}\,dr \,,
$$
and equals the (polynomial) degree of the rational map. Here, the
profile function $f(r)$ satisfies the boundary conditions $f(0) = \pi$ and
$f(\infty) = 0$. The energy can be written as
\begin{small}  
$$
E=4\pi \int_{0}^{\infty} \left(r^2 f'^2 + 2B\sin^2
f\left(f'^2+1\right)+ \mathcal{I}\,\frac{\sin^4 f}{r^2} 
+ 2m^2r^2\left(1-\cos f\right)\right) dr \,,
$$
\end{small}
where
$$
\mathcal{I} = \frac{1}{4\pi} \int
\left(\frac{1+|z|^2}{1+|R|^2}\left\vert\frac{dR }{ dz }\right\vert
\right)^4\frac{2i\,dz\,d\bar{z}}{(1+|z|^2)^2}\,.
$$
In order to minimise $E$ one first minimises $\mathcal{I}$ over all maps 
of degree $B$. The profile function $f$ is then found by solving the
Euler-Lagrange equation for $E$ with $B$ and $\mathcal{I}$ fixed.
For small $B$ the symmetry is sufficient to determine the 
relevant rational map. For larger $B$, the parameters have to be found
by minimizing ${\cal I}$ numerically. The rational maps which minimize
${\cal I}$ for $m=0$ have been determined numerically in
\cite{Battye:2001qn, Battye:2002wc} for all $B \le 40$. 
The rational map ansatz has proven to be very
successful for predicting energy and symmetries of Skyrmions (in
particular for $m=0$). 
Table \ref{T1} gives some examples of rational maps, and figure
\ref{F1} shows the corresponding level sets of constant energy
density.

\begin{center}
\begin{table}[ht]
\begin{center}
\begin{tabular}{c c c}
\hline\hline \\
$B$ & symmetry & $R(z)$ \\ [0.5ex] 
\hline \\
1 & $O(3)$  & $z$ \\ [2ex] 
2 & $D_{\infty h}$  & $z^2$ \\[2ex] 
3 & $T_d$  & $\frac{\sqrt{3}iz^2 - 1}{z^3 - \sqrt{3}iz}$ \\[2ex] 
4 & $O_h$  & $\frac{z^4 + 2\sqrt{3}iz^2 + 1}{z^4 - 2\sqrt{3}iz^2 + 1}$ \\[2ex] 
\hline
\end{tabular}
\caption{Rational maps for the minimal energy solutions for
  $B=1,\dots,4$. \label{T1}}
\end{center}
\end{table}
\end{center}

The restriction that that $R(z) = p(z)/q(z)$ is a quotient of two 
holomorphic polynomials can be relaxed. By allowing $p$ and 
$q$ to be functions of $z$ and ${\bar z}$ the rational map ansatz can be 
``improved''. This leads to best approximations to the numerical solutions 
for $B=3$ and $4$ known to date \cite{Houghton:2001fe}.  This approach
also provides insight into the singularity structure of Skyrmions. However, the
price to pay is that the energy now contains two functions which have
to be minimized (${\cal I}$ and ${\cal J}$). 
Another way to generalize the rational map ansatz is to ``deform the
sphere'', so that the energy  
density in localized around a squashed sphere. The baryon density remains 
the same, but the energy density picks up even more terms (work in 
progress). For yet another generalization of the rational map ansatz
see \cite{Ioannidou:2004nj}.

\subsection{Quantization of Skyrmions}

In the following, we describe how to quantize the Skyrme model, which
is a a scalar field theory, and obtain fermions following the approach
of Finkelstein and Rubinstein \cite{Finkelstein:1968hy}. Then we show
how to calculate the Finkelstein-Rubinstein constraints using the
rational map ansatz \cite{Krusch:2002by,Krusch:2005iq}. 

So, how do we quantize a scalar field theory and obtain fermions? The
key observation is that the configuration space of the Skyrme model is
topologically non-trivial, namely, 
\begin{equation}
\label{Pi1}
\pi_1 (Q_B) = \mathbb{Z}_2,
\end{equation}
where $Q_B$ denotes the space of Skyrme configurations with topological 
charge $B$. So, rather than defining the wavefunction on the
configuration space $Q_B$ we can define the wavefunctions $\psi$ on the
covering space of configuration space:
$$
\psi: CQ_B \rightarrow {\mathbb C}.
$$
Note that (\ref{Pi1}) implies that $CQ_B$ is a double cover.
In order to have fermionic quantization we have to impose the
following condition,
\begin{equation}
\label{loop}
\psi({\tilde q}_1) = - \psi({\tilde q}_2),
\end{equation}
where ${\tilde q}_1$ and ${\tilde q}_2 \in CQ_B$ project to the same point $q
\in Q_B$. In other words, ${\tilde q}_1$ and ${\tilde q}_2$ are
related by a non-contractible loop in configuration space. Such loops
naturally arise as symmetries of Skyrmions. For example, consider  a
rotation in space followed by a rotation in target space which leaves
a given Skyrme configuration invariant. Assuming that the wavefunction $\psi$
is localised at this Skyrme configuration leads to the following  
induced action of the $SO(3) \times SO(3)$ symmetry on $\psi:$
\begin{equation}
\label{symmetry}
\exp \left( -i \alpha\, {\bf n} \cdot {\bf L} \right)
\exp \left( -i \beta\, {\bf N} \cdot {\bf K} \right)
\psi({\tilde q}) = \chi_{FR} \psi({\tilde q}),
\end{equation}
$$
{\rm where}~\chi_{FR} = \left\{
\begin{array}{rl}
1 & {\rm if~the~induced~loop~is~contractible,} \\
-1& {\rm otherwise.}
\end{array}
\right.
$$
Here $\alpha,$ ${\bf n}$, $\beta$ and ${\bf N}$ are the angles and
axis of rotation in space and target space,
${\bf L}$ and ${\bf K}$ are the body-fixed angular momentum 
operators in space and target space, respectively. 
The key question is can we calculate $\chi_{FR} \in \pi_1(Q_B)$?
The answer is yes, and relies on the following theorem.

\begin{theorem}[S.K] 
The rational map ansatz induces a surjective 
homomorphism $\pi_1(Rat_B) \rightarrow \pi_1 (Q_B)$.
\end{theorem}

This theorem implies that if we can calculate the
Finkelstein-Rubinstein constraints for rational maps $R(z) \in Rat_B,$
where $Rat_B$ is the space of rational maps of degree $B$,  
then we also know the
constraints for the full configuration space $Q_B$ of the Skyrme
model. Fortunately, we can calculate the constraints for rational
maps, and hence obtain the Finkelstein-Rubinstein constraints 
$$
\chi_{FR} = (-1)^{\cal N} \quad {\rm where}~~{\cal N} = \frac{B}{2 \pi} 
\left(B \alpha - \beta \right),
$$
for the symmetry given by (\ref{symmetry}). 

This approach is very general, and it is worth emphazising that the
formula for the Finkelstein-Rubinstein constraints are exact, provided
the (numerically calculated) exact Skyrme configuration can be deformed
into a rational map Skyrmion while preserving the respective 
symmetries. However,
in order to calculate the wavefunctions, we have to make some
approximations. The simplest is the semiclassical collective
coordinate quantization which we will describe in the following. 

Starting from a static Skyrme configuration $U_0({\bf x})$ we can
generate a set of Skyrme configurations $U({\bf x})$ which have the
same energy as $U_0({\bf x})$ via
$$
U({\bf x}) = A\ U_0\left(D\left(A^\prime\right) {\bf x}\right)A^\dagger
$$
where $A$ and $A^\prime$ are constant $SU(2)$ matrices and
$D(A^\prime)$ is the associated $SO(3)$ rotation. Inserting this ansatz
into the Skyrme Lagrangian (\ref{Lagrangian}) we obtain the following
kinetic energy
\begin{equation}
\label{T_eff}
T =\frac{1}{2}a_i U_{ij} a_j - a_i W_{ij} b_j + \frac{1}{2}b_i V_{ij}
b_j \,, 
\end{equation}
where $a_k = -i\hbox{Tr}\,\left(\tau_k A^\dagger {\dot A}\right)$ 
and $b_k= -i\hbox{Tr}\,\left(\tau_k {\dot
  {A^\prime}}{A^\prime}^\dagger\right)$.  
Here $U_{ij},$ $V_{ij}$ and $W_{ij}$ are
integrals involving $U_0$. The conjugate momenta corresponding to $b_i$
and $a_i$ are the body-fixed spin and isospin angular momenta $L_i$
and $K_i$:  
\begin{eqnarray*}
L_i &=& -W_{ij}^T a_j + V_{ij}b_j\,,\\
K_i &=& U_{ij} a_j - W_{ij}b_j\,.
\end{eqnarray*}
We denote the space-fixed spin and isospin angular momenta by 
$J_i$ and $I_i$ respectively. Note ${\bf J}^2 = {\bf L}^2$ and ${\bf I}^2 
= {\bf K}^2$. We now regard $L_i$, $K_i$, $J_i$ and $I_i$ as quantum 
operators, each individually satisfying the $\mathfrak{su}(2)$ commutation 
relations\footnote{For a discussion of the more
  subtle points about body-fixed and space-fixed angular momenta in
  this context see for example \cite{Braaten:1988cc,Krusch:2005bn}}.

The idea is now to find the lowest values of spin and isospin which 
are compatible with all the Finkelstein-Rubinstein constraints, then 
calculate the energy of these states. The 
Finkelstein-Rubinstein constraints arise from the symmetries of 
the classical minimal energy configuration. A basis for the
wavefunctions is given by  
$|J,J_3,L_3\rangle
\otimes |I,I_3,K_3\rangle$. 
Via Legendre transform, 
the kinetic energy $T$ can be expressed in terms of angular momentum 
operators. The matrices $U_{ij}$, $V_{ij}$ and $W_{ij}$ can be
evaluated using  
\begin{small}
$$
\Sigma_{ij} = 2\int \sin^2 f  
\frac{C_{\Sigma_{ij}}}{(1+|R|^2)^2}\left(1+f'^2+\frac{\sin^2 
f}{r^2}\left(\frac{1+|z|^2}{1+|R|^2}\left|\frac{dR}{dz}\right|\right)^2 
\right)\,d^3x\,,   
$$
\end{small}
where $\Sigma = (U,V,W)$ and $C_{\Sigma_{ij}}$ only depends on angular 
variables \cite{Manko:2007pr}. A similar expression was derived in 
\cite{Kopeliovich:2001yg}.

\subsection{Results and Challenges}

The approach in the previous section has been carried out quite
successfully by various
authors. The groundstates and possible excited states have been calculated 
using Finkelstein-Rubinstein constraints for $B \le 22$,
\cite{Irwin:1998bs,Krusch:2002by}, leading to satisfactory results for
even $B$, but the predictions for odd $B$ are not as good. 
There are also some general results for
even-even and odd-odd nuclei, which agree with experiment
\cite{Krusch:2005iq}. Manko, Manton and Wood \cite{Manko:2007pr}
calculated the energy levels for  $1 \leq B \leq 8$. The results
correspond well to experiment, and energies and spins of a few as yet
unobserved states have been predicted.  

However, the quantitative results are not very accurate. For odd $B=5$
and $7$, the correct ground states can only be calculated by deforming
the minimal energy solutions significantly, \cite{Manko:2006dr}. For
$B=10$, $18$ and $22$, the approach gives the wrong groundstates. 

Despite these seemingly disappointing results there is cause for
optimism. There are various ways in the above approximations can be
improved. The zero-mode quantization is the simplest approximation to the 
quantum states (also known as rigid rotator approximation). A more
sophisticated quantization can lead to better results
\cite{Houghton:2005iu, Leese:1995hb, Acus:2006bp}

There is also an ongoing discussion of how best to fix the three physical 
parameters  ($f_\pi$, $e$ and $m$) in the Skyrme model.
The original, and most widely used, set of Skyrme
parameters has been proposed in \cite{Adkins:1983ya,Adkins:1984hy},
by matching to the proton and the Delta mass. However, it has been
shown that this matching condition can be considered to be an artifact
of the rigid rotator approximation, \cite{Battye:2005nx}. The studies
in \cite{Battye:2006na} suggest that the effective pion mass $m$
should have the rather large value of $m \approx 1$. In
\cite{Kopeliovich:2004pd}, a 30\% lower value of the Skyrme
parameter was suggested in order to match a large range of nuclei masses to
experimental data. A similar conclusion was reached in
\cite{Manton:2006tq} by considering the electromagnetic properties of the
quantized $B=6$ Skyrmion, describing $_3^6$Li.

Another recent development is the realization that 
the pion mass which occurs in the last term in (\ref{Lagrangian}) 
has a larger influence than expected. At the physical value of the pion
mass, shell-like configurations become unstable to squashing
\cite{Battye:2004rw}. While the minimals for small $B < 8$ hardly
change, many new (local) minimal energy configurations have been found
for larger $B$, see \cite{Battye:2006tb, Battye:2006na,
  Houghton:2006ti}. These new configurations are no longer shell-like,
but often seem to contain several $B=4$ cubes. The relation to the
phenomenological alpha particle model is discussed in
\cite{Battye:2006na}. There is hope that the Skyrme model will one day
be able to describe the low energy excited states for example for
$B=12$ or $B=16$ as rotational bands related to different local minima
in the same way as in the alpha particle model. In order to
achieve this aim, we need a better understanding of the classical
solutions, possibly including saddle point solutions in the Skyrme
model \cite{Krusch:2004uf}. 

Finally, we have to take into account that Skyrmions
deform when they are spinning. This can lead to changes in the
symmetries of minimal energy solutions, and different symmetries can
lead to different allowed quantum states!

\section{Relative equilibria}
\label{Rel}

The zero mode quantization has been quite successful. However, 
it is clear that Skyrmions are deforming when they are spinning, and this 
effect needs to be taken into account. The theory of relative
equilibria was developed for rotating molecules. It gives strong
theorems as to what kind of symmetry bifurcations can occur. 
Numerical simulation of spinning Skyrmions is very time-consuming. 
Therefore, a good understanding of what kind of behaviour is to be 
expected is very valuable.

\subsection{Theoretical Background}

Let ${\cal C}$ be a smooth manifold with a smooth action of a 
finite dimensional Lie group $G$. Let $<.,.>_q$ denote the $G$-invariant 
inner product on $T_q{\cal C}$. Consider the $G$-invariant Lagrangian
$$
L = \tfrac{1}{2} <{\dot q}, {\dot q} >_q - V(q),
$$
where $V(q)$ is also $G$-invariant. Let $q(t) = \exp(t \xi)
r(t)$. This induces the Lagrangian $L_\xi$ 
$$
L_\xi = \tfrac{1}{2} <{\dot r}, {\dot r}> -
<{\dot r}, \xi \cdot r> + \tfrac{1}{2} < \xi \cdot r, \xi \cdot r > - 
V(r),
$$
where 
$$\xi \cdot r = \frac{d}{dt} \left(\exp(t \xi) r\right)_{t=0}.$$
An equilibrium point of $L_\xi$ is called a relative equilibrium of 
$L$. A good choice of coordinates are the {\em slice coordinates,} splitting 
the dynamics into components along the group orbit $G \cdot q_0$ and 
transverse to it where $q_0$ is any point in ${\cal C}$. In a 
neighbourhood of $q_0$, we can write $q = g \cdot s$ where $g \in G$ and 
$s$  is in the ``slice'' $S$ transverse to the group action. 
For the tangent vectors, we obtain
$$g^{-1} {\dot q} = \xi \cdot s + {\dot s},$$
where $\xi = g^{-1}{\dot g}$. These coordinates have the advantage 
that rotational and vibrational modes decouple at the point $q_0$.

\begin{theorem}
\begin{enumerate}
\item In a neighbourhood of an equilibrium point $q_0$ the corresponding 
Hamiltonian $H$ can be 
expressed as
$$
H = h(\mu) + Q(s,\sigma),
$$ 
where $\mu$ and $\sigma$ are the momenta conjugate to $\xi$ and $s$, 
respectively.
\item The function $h(\mu)$ is even and its Taylor series at $0 \in 
{\mathfrak g}^*$ is
\begin{small}
$$
h(\mu) = V(q_0) + \tfrac{1}{2} \mu^T {\bf I}(q_0)^{-1} \mu
- \tfrac{1}{16} V_2^{-1}\left(
\mu^T {\bf I}_s^{-1}(q_0) \mu, \mu^T
{\bf I}_s^{-1}(q_0) \mu
\right) + O(\mu^6),
$$
\end{small}
where ${\bf I}$ is the locked inertia tensor, and $V_2$ and ${\bf 
I}_s^{-1}$ are certain derivatives with respect to $s$.
\end{enumerate}
\end{theorem}

This theorem gives us a way to systematically explore relative
equilibria in the Skyrme model. Here, $\mu$ corresponds to the angular
momenta ${\bf K}$ and ${\bf L}$, and $s$ parametrizes the
vibrations/deformations, which we have ignored so far. A better
understanding of the classical behaviour will help us to derive a more
sophisticated quantization of Skyrmions. Note that the zero-mode
quantization corresponds to only considering the quadratic
approximation to $h(\mu)$. Symmetries place further restrictions on
$h(\mu)$ and enable us to make more general statements about possible
bifurcation patterns.

\section{Results and Outlook}

In this talk, we described how to use the Skyrme model to calculate
ground states and excited states of atomic nuclei. We also discussed
various effects that have to be taken into account. One important
effect is that Skyrmions deform when they are spinning. The
theory of relative equilibria, which has been very useful for
understanding molecular spectra \cite{Montaldi:1999, Kozin:1999,
  Kozin:2000}, will help us to 
find the answer to various questions. 

The first set of questions is about classical behaviour. 
We want to understand the families of relative equilibria which
bifurcate from a given equilibrium point. Furthermore, we want to
understand the symmetries of various bifurcations given the symmetry
of the equilibrium points and the locked inertia tensor (which is
related to the matrices $U_{ij}$, $V_{ij}$ and $W_{ij}$ in equation
(\ref{T_eff})). We are planning to calculate these symmetry
bifurcations explicitly for various baryon numbers. Using a
generalization of the rational map ansatz it might be possible to
evaluate higher order terms in the function $h(\mu)$, which contains a
lot of information about the bifurcations. This approach should at
least enable us to identify interesting configurations together with
the most appropriate rotation axes in space and target space. The
corresponding relative equilibria could then be calculated
numerically. 
For higher baryon numbers there is often a multitude of different local
minima. 
The knowledge of how for a given $B$ the bifurcation pattern of
these various families evolves as the angular velocity increases will
give us a better insight into which of these minima play the most
important role. 

Finally, there is an extension of the theory of relative equilibria to
quantum mechanical problems which can also be applied to the Skyrme
model. Hopefully, this theory, together with the insight we gained from
the classical problem, will allow us to calculate ground state and
excited states of light atomic nuclei with more confidence, including
those with half-integer spin.

\section*{Acknowledgments}

S. K. would like to thank the organizers of Quarks-2008 for a very
interesting conference. S. K. is grateful to N. S. Manton, J. M. Speight
and S. W. Wood for fruitful discussions. The work of M. R. was supported by 
a Leverhulme Trust Research Fellowship.  

\bibliographystyle{/home/sk211/bib/myordered}
\renewcommand{\baselinestretch}{1}
\addcontentsline{toc}{section}{Bibliography}
%

\begin{small}

\end{small}
\label{lastref}

\end{document}